# Crystal Growth and Characterization of HgBa$_2$Ca$_2$Cu$_3$O$_{8+\delta}$ Superconductor with the Highest Critical Temperature at Ambient Pressure

Bastien Loret[1,5], Anne Forget[1], Jean-Baptiste Moussy[1], Sylvie Poissonnet[2], Patrick Bonnaillie[2], Gaston Collin[3], Pierre Thuéry[4], Alain Sacuto[5] and Dorothée Colson[1]*

1 SPEC, CEA, CNRS-UMR 3680, Université Paris-Saclay, Gif sur Yvette Cedex 91191, France
2 SRMP, DMN, CEA, Université Paris-Saclay, Gif sur Yvette Cedex 91191, France
3 LPS, C.N.R.S. UMR 8502, Université Paris-Sud, Orsay 91405, France
4 NIMBE, CEA, CNRS, Université Paris-Saclay, Gif sur Yvette Cedex 91191, France
5 Laboratoire Matériaux et Phénomènes Quantiques, 10, rue A. Domon et L. Duquet, PARIS Cedex 13 75205, France

*Supporting Information Placeholder*

**ABSTRACT:** We report an original procedure for the elaboration of very high quality single crystals of superconducting HgBa$_2$Ca$_2$Cu$_3$O$_{8+\delta}$ mercury cuprates. These single crystals are unique with very high quality surface paving the way for spectroscopic, transport and thermodynamic probes in order to understand the hole-doped cuprate phase diagram. The superconductivity transition width of about 2 K indicates they are homogeneous. Annealing allows to optimize $T_c$ up to $T_c^{max}$ = 133 K. We show for the first time that with adequate heat treatment, Hg-1223 can be largely under-doped and its doping level controlled. Importantly, the crystal structure was studied in detail by single crystal X-ray diffraction, and we have identified the signature of the under-doping by a detailed sample characterization and micro-Raman spectroscopy measurements.

The discovery of the cuprate superconductors triggered a very dynamic search for new compounds, while the identification and understanding of their physical properties are still a true challenge because of their sensitivity to chemical substitution or modification of their non-stoichiometric oxygen content. After 30 years of research, a number of questions remain regarding in particular the electron pairing mechanism involved, the high critical temperature $T_c$ and its dome-like shape dependence on doping, the relationship between the structure and the electronic properties, and the effect of pressure on superconductivity. Yet, this understanding is crucial in order to reach higher $T_c$ values that would allow the development of considerable technological applications. Mercury cuprates HgBa$_2$Ca$_{n-1}$Cu$_n$O$_{2n+2+\delta}$, where $n$ is the number of CuO$_2$ layers, might be good candidates to meet this challenge. Since their discovery in 1993, they still hold the record value of $T_c$ in the cuprates family, 133 K (160 K under 30 GPa) for HgBa$_2$Ca$_2$Cu$_3$O$_{8+\delta}$ (Hg-1223).[1–4] Unfortunately very few studies have been done on this compound because no single crystals were available, although mercury compounds can be considered as model structures for understanding the mechanisms of high $T_c$ superconductivity, given their high tetragonal symmetry (Figure 1) and their record high $T_c$ values.

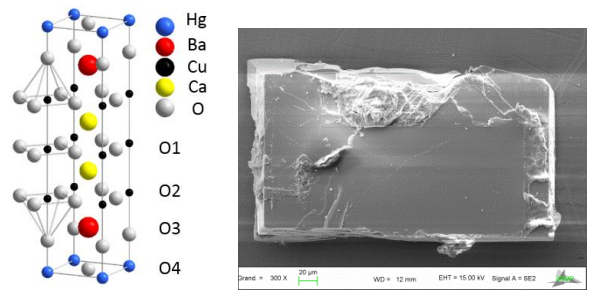

**Figure 1.** Schematic representation of the tetragonal HgBa$_2$Ca$_2$Cu$_3$O$_{8+\delta}$ crystal structure (left) and SEM view of a single crystal (right). The surfaces of the crystals are clean and the crystallographic axes are easy to identify ($a$-axis along the edges, $c$-axis perpendicular to the platelet).

Here we report a new method to grow single crystals of Hg-1223 cuprate with self-flux growth technique. The crystals thus synthesized show high surface quality for spectroscopic measurements and display a superconductivity transition width of about 2 K which suggests they are homogeneous. Annealing enable to optimize $T_c$ up to the $T_c^{max}$ of 133 K. We show for the first time that with adequate heat treatment, we can largely under-dope the Hg-1223 compound and control its doping level. Importantly, we have identified the signature of the under-doping by a detailed sample characterization and micro-Raman spectroscopy measurements. Note that the crystals elaborated by the method described here have already enabled recent Raman studies.[5]

Even if the mercury compounds are very attractive, they are however complex oxides and their synthesis remains a challenge. The first impediment in preparing the mercury-based cuprates lies in the toxicity of HgO and the volatility of its decomposition product, namely Hg, which occurs near 450 °C in air, before the mercury-containing superconducting phases are formed. Therefore the first synthesis route used to produce pure phases was via a solid state reaction in a sealed tube,[1-3] but the pure calcium-containing phase ($n \geq 2$) was obtained only with difficulty; high pressure techniques have been shown to prevent HgO decomposition and to facilitate the formation of the desired products.[6-10] These very useful techniques have shown their efficiency to obtain the phases with a high $n$ number of $CuO_2$ layers ($n \geq 4$).[11-15] Because of the complexity of its compounds chemistry, considerable difficulties remain in growing single crystals of mercury cuprates and separating them from the flux.[11,16,17,18,19] Indeed the compounds undergo incongruous melting and decompose at low oxygen pressure, having low thermal and chemical stability.

All these chemical obstacles explain the limited number of crystal growth reports on pure or substituted Hg-1223.[16,17,18] Here we show that our method allows the synthesis of large pure crystals and the easy separation of the sample from the flux, thus yielding crystals with an extremely high quality and clean surface. This method has the advantage to be less expensive, less delicate and more simple than the technique using gold foil.[17]

A low melting region in the pseudo-ternary BaO–CaO–CuO diagram was determined, and crystal growth was achieved in a melt with an excess of BaO and CuO (see Supporting Information). In our experimental conditions, the interesting domain is bounded by $BaCuO_2$, CaO and CuO (Figure S1, Supporting Information). The main compositions along the $Ba_2CuO_3$–$CaCuO_2$ and Hg1223–CuO lines have been investigated. In order to determine a suitable growth temperature for the compositions labelled A to G, samples were heated to temperatures between 800 and 985 °C and then cooled down at 5°C/h. The experiments have revealed that a step at 700 °C where the Hg-1223 is formed greatly improves the reactivity and ensures a complete reaction by minimizing the soaking time.

The compositions most favourable to crystal growth were B and F, where 12 mol% HgO and respectively 58:30 and 62:26 mol % BaO:CuO were mixed. With the compositions richer in CuO (I and G), mostly $BaCuO_2$ and CuO crystals were obtained. After the growth stage, most of the crystals are found at the bottom of the crucible in the frozen flux and they need to be mechanically separated from it. The surfaces of the crystals are extremely clean and show no trace of remaining flux attached to them, which is probably due to the low viscosity of the melt allowing better separation. Figure 1 shows one typical, platelet-shaped crystal, with well-developed {001} faces (0.4 × 0.5 × 0.3 mm³).

A remarkable point is that the crystallographic axes of the crystalline platelets are easily identifiable (*a*-axis along the edges, *c*-axis perpendicular to the platelet largest face). The chemical analysis of as-grown crystals, performed with a scanning electron microscope equipped with an electron microprobe, revealed a slight under-stoichiometry of ~0.94 in mercury, the mercury content being homogeneous in each crystal. Globally there is good agreement with the expected formula and no flux component was detected.

The crystal structure was studied by single crystal X-ray diffraction (see Supporting Information for details). Atomic coordinates are given in Table S1 (Supporting Information), and crystal data and refinement details in Table S2 (Supporting Information). In all cases, the occupancy parameter of the mercury cation was refined, as well as that of the oxygen atom O4 at (1/2 1/2 0). The displacement parameter of the latter atom, which corresponds to a low electronic density located close to several heavy atoms, is strongly unstable and cannot be freely refined anisotropically; it was instead constrained to be isotropic with a value of twice the equivalent isotropic displacement parameter of atom O3 (so as to account for the different environments of these two atoms); the choice of this multiplicative factor has however no very marked effect and it was checked that its variation in the range 1.2–3.0 has either no effect (for the samples with the smallest occupancies), or induces variations in the occupancy factor at most equal to half the error bar (for the samples with the highest occupancies). An oxygen atom introduced within the material transfers two carriers into the copper-oxygen ($CuO_2$) planes, and a universal relation between $T_c/T_c^{max}$ and the number of carriers in $CuO_2$ planes has been found.[20] The carriers concentration and associated additional oxygen content determined in this way are in satisfying agreement with the values from diffraction data, when the large standard deviations on the latter are taken into account (Table S2).

All previous studies but one,[21] whether carried out by neutron powder diffraction or X-ray diffraction on crystallites, result in incomplete occupancy at the mercury site,[22-25] the origin of which was discussed for a long time, with numerous models invoking the presence of carbonate or copper ions, or vacancies at the mercury site having been proposed.[16,23–27] However, thermal post-growth treatments at higher temperatures systematically increase the mercury occupancy: $n$(Hg) ~0.93 for annealed crystals at 400 °C under vacuum to be compared to $n$(Hg) ~0.90 for as-grown or oxygenated crystals at 300 °C. Moreover this effect is reversible: a crystal first treated at 400 °C under vacuum, then at 300 °C in oxygen flux exhibits the same $n$(Hg) ~0.90 as crystals as-grown or simply oxygenated at 300 °C. These results are not consistent with the model invoking the substitution of copper on the mercury site which should be stable under annealing conditions. An interpretation is suggested by our new measurements: it is as if part of mercury was dispersed on a more general position than the special position 1*a* of *P*4/*mmm*, and the amount of this disordered mercury fraction depends, as expected, on the temperature and atmosphere of annealing, a part of it being reintroduced on the regular mercury site at 400 °C. This last effect is unfortunately limited because the Hg1223 phase decomposes upon annealing at higher temperatures. But the question of the location of the 'missing' mercury, either in general position or as a quasi-liquid distribution, remains open and difficult to answer by usual Fourier transform techniques.

Magnetic measurements show that the as-grown samples are under-doped with a $T_c$ around 105 K (AG105) and a fairly sharp transition, as shown in Figure 2. In order to increase $T_c$ to $T_c^{max}$, an oxidant annealing procedure under molecular oxygen flux at 325 °C was developed. The crystals were placed in advance in a mixture of HgO, BaO and CuO oxides to prevent a departure of mercury from the samples during the 10 days heat treatment.

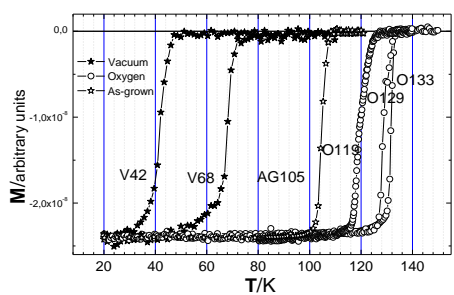

**Figure 2.** Magnetic susceptibility measurements of as-grown, oxygenated and under vacuum-annealed crystals of $HgBa_2Ca_2Cu_3O_{8+\delta}$.

After annealing, $T_c$ increases up to 133 K (O129 and O133) with a narrow transition width of 5 K confirming the good homogeneity of the oxygen content in the sample. In order to access the under-doped region, we also developed a procedure of annealing under vacuum. As-grown crystals have been annealed under a vacuum of $5.0\ 10^{-7}$ mbar at 350 °C during 10 days, or at 425 °C during 5 days. These thermal treatments decrease considerably $T_c$ from 105 K to respectively 68 K (V68) and 42 K (V42), as shown in Figure 2. This low $T_c$ value of 42 K has never been reached so far in Hg-1223.

These crystals have been characterized by micro Raman spectroscopy which reveals for the first time the specific Raman features of the as-grown, optimally doped and under-doped Hg-1223 compounds. The Raman spectrum of a typical as-grown single crystal (AG105) obtained in the $x(zz)x$ configuration (see Supporting Information) is displayed in Figure 3.

This Raman spectrum (black curve) exhibits a phonon peak at 591 cm$^{-1}$ assigned to the in-phase motion (along the $c$-axis) of the apical oxygen atoms (O3) (Figure 1).[26,27] This peak assumes an asymmetric shape with a maximum at 591 cm$^{-1}$ and a shoulder at 582 cm$^{-1}$. Additional weaker intensity features at 560 cm$^{-1}$ and 539 cm$^{-1}$ can also be detected. The Raman spectrum of the optimally doped (O133) single crystal (red curve) is obtained from an as-grown single crystal after oxygen annealing. We can notice that the integrated Raman intensity of the apical oxygen (O3) mode decreases by more than half compared to the as-grown crystal and also shifts to low frequency with a maximum centered around 578 cm$^{-1}$. On the contrary, the Raman spectrum (green curve) of the under-doped single crystal got from an as-grown crystal after vacuum annealing displayed an increase of the integrated Raman intensity of the apical oxygen (O3) mode and it shifts to high energy up to 596 cm$^{-1}$. A second well defined peak at 539 cm$^{-1}$ is also detected. It corresponds to the in-phase motion (along the $c$-axis) of the oxygen (O2) atoms related to the two $CuO_2$ planes located on each side of the $CuO_2$ symmetry plane (Figure 1). Therefore, Raman spectroscopy allows us to clearly distinguish between the under-doped and the optimally doped single crystals and confirms the efficiency of the annealing procedure.

This communication demonstrates that the flux growth technique is effective for the growth of single crystals of $HgBa_2Ca_2Cu_3O_{8+\delta}$ with well-controlled oxygen content and very clean surfaces ideal for optical measurements. This original procedure allows the growth of very high quality single crystals of mercury cuprates with the highest critical temperature (133 K) at atmospheric pressure. The crystals have been characterized by different techniques demonstrating their high quality. Our annealing method in vacuum has enabled to prepare, for the first time, Hg-1223 crystals at different doping levels in the under-doped regime. Raman scattering confirms the quality of the crystals and our ability to change the doping level. This work is an important step forward, which will allow a deeper understanding of the hole-doped cuprate phase diagram, in particular the true nature of the mysterious pseudogap phase and its relationship to the superconducting state. More generally, this flux growth technique can be applied to other compounds for which there exists no phase diagram and/or the growth in air is forbidden because of volatile components.

## ASSOCIATED CONTENT

**Supporting Information**

The Supporting Information is available free of charge on the ACS Publications website at DOI:.

    Experimental details, Tables and additional Figure. (PDF)

    X-ray crystallographic information. (CIF)

## AUTHOR INFORMATION
**Corresponding Authors**
*E-mail: dorothee.colson@cea.fr (D. C.)

**Notes**
The authors declare no competing financial interest.

## ACKNOWLEDGMENT

Financial support from the DIM OxyMORE for PhD thesis of Bastien Loret is acknowledged.

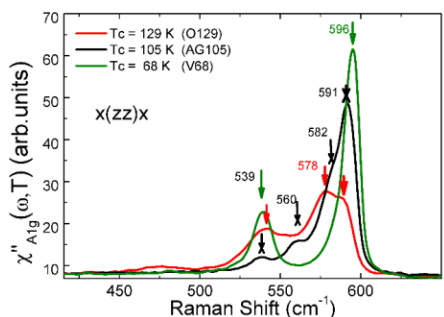

**Figure 3**. Raman spectra of as-grown, optimally doped and under-doped Hg-1223 single crystals. The phonon lines are related to the oxygen vibrational modes which are strongly affected by the doping level variation.

# Crystal Growth and Characterization of HgBa$_2$Ca$_2$Cu$_3$O$_{8+\delta}$ Superconductors with the Highest Critical Temperature at Ambient Pressure

Bastien Loret, Anne Forget, Jean-Baptiste Moussy, Sylvie Poissonnet, Patrick Bonnaillie, Gaston Collin, Pierre Thuéry, Alain Sacuto, Dorothée Colson[*]

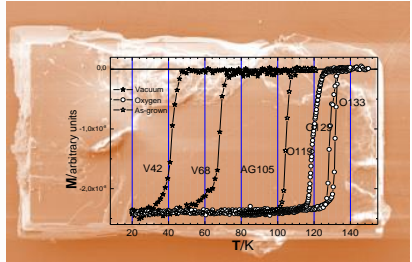

An original procedure for the elaboration of very high quality single crystals of superconducting HgBa$_2$Ca$_2$Cu$_3$O$_{8+\delta}$ mercury cuprates is described, which allows a fine tuning of the oxygen content and will enable to probe, in a large range of composition, the phase diagram of these compounds displaying the highest critical temperature at ambient pressure.